\documentstyle[prb,aps,psfig,floats]{revtex}
\psfigurepath{/home/zeke1/users/radtke/papers/dqw/asym}

\begin{document}

\twocolumn[
\hsize\textwidth\columnwidth\hsize\csname@twocolumnfalse\endcsname

\title{Mode mixing in Antiferromagnetically Correlated Double Quantum Wells}
\author{R. J. Radtke}
\address{Division of Engineering and Applied Sciences,\\
Harvard University, Cambridge, Massachusetts~~02138}
\author{S. Das Sarma}
\address{Condensed Matter Theory Group, Department of Physics,\\
University of Maryland, College Park, Maryland~~20742-4111}
\author{A. H. MacDonald}
\address{Department of Physics, Indiana University,
Bloomington, Indiana~~47405}
\date{Received by Phys. Rev. B on 7 July 1997}
\maketitle

\begin{abstract}

We examine the robustness of a recently predicted
exchange-induced zero-field magnetic phase in semiconductor
double quantum wells in which each well is spin-polarized
and the polarization vectors are antiparallel.
Magnetic instabilities are a general feature of Coulombic
double quantum well systems at low densities.  We argue
that this antiferromagnetic phase is stabilized relative to
ferromagnetic ones by an effective superexchange interaction
between the wells.
Detailed self-consistent Hartree-Fock calculations using a
point-contact model for the interaction show that the
antiferromagnetic phase survives
intra-subband repulsion matrix elements 
neglected in earlier work in a large portion of the model's parameter
space.  We also examine the role of asymmetry due to biasing or to
differences in the widths of the two quantum wells.  
The asymmetry creates a mode coupling between
the intra- and inter-subband collective spin-density excitations
(SDEs) which changes the Raman spectroscopy signature of the
phase transition from a complete softening of the inter-subband SDE
to a cusp as the density is tuned through the transition.
This cusp may be detectable in inelastic light scattering
experiments in samples of sufficient quality at low enough
temperatures and densities.
\\
\end{abstract}


]


\section{INTRODUCTION}
\label{sec:introduction}

The subject of exchange-correlation-induced phase
transitions has proven to be a rich field of research
which has revealed many intriguing phenomena.
Normal Fermi-liquid-state instabilities occur when the kinetic energy
of the particles in a quantum system is of the same order or
smaller than the inter-particle exchange and correlation
energies.  The instabilities lead to a variety of new electronic
states.  In Coulombic systems, this situation obtains at low densities
and instabilities are more likely in systems of 
reduced dimensionality or, especially in two dimensions, in an applied
magnetic field. 
Perhaps the best known examples of such new states occur
in the in the fractional-quantum-Hall, strong-field regime of two-dimensional
electronic systems.\cite{fqhe}

The interest in exchange-correlation-induced phase transitions
can be traced in part to the availability of high-quality
semiconductor quantum wells, quantum wires, and superlattices.
These artificial structures may be fabricated with
remarkable precision and quality and possess electron
densities that can be varied over a large range through a
combination of modulation doping and judicious gating.
The one- or two-dimensional nature of the resulting
electronic systems, and the low densities realizable
in devices of this kind, make them ideal for studies of
interaction-driven physics.
Another feature of these devices of importance to the
current work is the extra degree of freedom available when
multiple layers are present, as in multiple quantum wells or
superlattices.
This new degree of freedom allows transitions to states
with order not only in the intra-layer electronic degrees of
freedom but also in diagonal or off diagonal inter-layer 
charge \cite{MacDonald,Ruden,nei,Varma,Katayama,Ying,Patel,Conti,Zheng,Wen-Zee_double,ezawa,iu_double}
or spin\cite{Zheng,das-tam,rad-das,Pinczuk,Zheng-Rad-Das,Plaut}
observables. 

In particular, a great deal of attention has recently focussed
on the possibility of inter-layer spin ordering in wide
single or double quantum wells.
This attention is motivated by studies of quantum
well structures in which the lowest two subbands are well
separated in energy from the higher subbands and the density
is sufficiently low so that only these two subbands are
occupied.
In these structures, an earlier theoretical calculation of the
collective spin-density excitations (SDEs) in the
absence of a magnetic field showed a complete softening of
the inter-subband SDE in a range of densities around that
at which the second subband begins to populate.\cite{das-tam}
Subsequent analysis indicated that this softening
corresponded to a phase transition to a state in which each
well in the double quantum well (or the electron gases on
each side of a wide single quantum well) was spin
polarized with the polarization vectors
antiparallel; i.e., a transition to an antiferromagnetic
order in the well spin densities.\cite{rad-das}
Further work predicted that a similar transition to a canted
antiferromagnetic phase should occur in the presence of a
magnetic field at $\nu = 2$,\cite{Zheng-Rad-Das} and this
transition seems to have been observed
experimentally.\cite{Pinczuk}
However, the predicted transition in the zero-field case has
not yet been observed.\cite{Plaut}

Several possibilities exist which may explain the absence
of the zero-field antiferromagnetic phase in these
experiments.
The calculations predicting this phase\cite{rad-das} are
based on a mean-field treatment of the interacting system,
which is known to overestimate the densities and temperatures
at which such symmetry-breaking transitions occur.
In addition, electronic scattering by disorder or impurities
generally has a detrimental effect on correlation-induced
phases.
Both of these difficulties are exacerbated by the low
dimensionality of the double quantum well system considered.
Thus, the zero-field antiferromagnetic phase may exist but
may not have been observed due to measurements made at
temperatures which are too large in samples of insufficiently
high quality.
The fact that the $\nu = 2$ transition is observed\cite{Pinczuk}
as predicted by the mean-field calculations\cite{Zheng-Rad-Das}
does not contradict this point of view, since the magnetic
field completely quenches the kinetic energy and makes the
mean-field theory a controlled approximation to the interacting
system.

Alternative explanations for the absence of the zero-field 
phase lie in the structure of the theory itself,\cite{rad-das}
which was derived to explore the qualitative features
of the antiferromagnetic phase without considering
several confounding effects which may nonetheless be
important.  In particular, the previous calculation\cite{rad-das} 
did not account for interactions between 
electrons in the same subband, which should be of the 
same size or stronger than the interactions between electrons 
in different subbands which were included.   This omission can become
especially important when the intersubband excitation softens.
These interactions could introduce ferromagnetic
or charge-ordered phases into the model which are not probed
by the current experiments.
More seriously, the slight asymmetry present in any
realistic double quantum well structure will couple the
intra- and inter-subband SDEs, potentially preventing the
latter from softening.
As this softening was expected to be a hallmark of the
antiferromagnetic phase transition,\cite{das-tam,rad-das}
it seemed reasonable to hypothesize that any asymmetry in
the structure might suppress the antiferromagnetic phase entirely.

In this paper, we address the question of the robustness of
the zero-field antiferromagnetic phase in the presence of
intra-subband interactions and asymmetry in the double
quantum well structure.
First, we argue on general grounds that the antiferromagnetic phase
is a direct consequence of the importance of the intra-well
exchange interaction at low densities and is stabilized by the
inter-well hopping, which leads to an effective
superexchange interaction.
Thus, this phase should obtain in a suitably constructed
heterostructure.
Second, we extend the self-consistent Hartree-Fock
calculations of earlier work\cite{rad-das} to include both
intra- and inter-subband matrix elements of the model
interaction and the effects of an asymmetric double
quantum well.
These calculations demonstrate that, while the intra-subband
interaction does introduce ferromagnetic phases and 
asymmetry does reduce the region of the phase diagram
occupied by the antiferromagnetic phase, the
antiferromagnetic phase does not disappear.
By examining the collective mode spectrum in the asymmetric
structure, we also find that the inter-subband SDE does not
soften due to the mode coupling between the intra- and
inter-subband SDEs.
The antiferromagnetic transition nevertheless occurs as a
result of the collapse of the {\it intra}-subband SDE which,
through the mode-coupling, has a strongly antiferromagnetic character.

The outline of this paper is as follows.
In the next Section, we employ a simple model for two weakly
coupled two-dimensional electron gases to examine the energetics
of the antiferromagnetic transition.
Sec.~\ref{sec:formalism} contains the formalism for the
extended self-consistent Hartree-Fock theory used in the
remainder of the paper.
This formalism includes asymmetry and all matrix elements
of the interaction and is used to compute both the ground
state and collective mode properties in what follows.
In Sec.~\ref{sec:results}, we present the results of our
computations for the matrix elements, phase diagram, and
collective modes in this model and discuss their
implications.
Finally, in Sec.~\ref{sec:conclusion}, we summarize our
results and conclude.

\section{ORIGIN AND STABILITY OF THE ANTIFERROMAGNETIC PHASE}
\label{sec:origin}

In this Section, we examine a simple model of a
double quantum well in order to extract the basic
physics underlying the zero-field antiferromagnetic phase.
To that end, consider two two-dimensional electron gases
separated by a barrier constructed so that the interaction
between them is negligible.
This would be the case if the two-dimensional layers were
widely separated or the barrier were very high and if the
dielectric constant of the barrier were large.
Suppose the electrons in each layer move freely 
except for a Hubbard-like point-contact interaction
$V({\bf r}) = V_0 \delta({\bf r})$, as used in textbook
treatments of itinerant magnetism.\cite{Doniach}

For equal charge density in the two layers,
the Hartree-Fock energy $E_{\rm HF}$ of the
two-layer system can be derived following
Refs.~\onlinecite{MacDonald,Ruden,Zheng} and may be expressed 
in the form 
\begin{equation}
\frac{E_{\rm HF}}{A} = 
  \frac{n^2}{8 N_0}\,(1-N_0 V_0)
  \left( \frac{1 + m_1^2}{2} + \frac{1 + m_2^2}{2} \right)
  + \frac{1}{4} V_0 n^2,
\label{eq:Ehf}
\end{equation}
where $n$ is the total electronic density in both layers,
$N_0$ is the single-spin, two-dimensional density of states,
and
\begin{equation}
m_i =
  \frac{n_{i\uparrow}-n_{i\downarrow}}
       {n_{i\uparrow}+n_{i\downarrow}}
\label{eq:m}
\end{equation}
is the relative spin polarization in layer $i = 1,2$ with
partial spin-dependent densities $n_{i\sigma}$.
In this equation, the first term represents the contribution
to the energy from combined kinetic and exchange effects,
and the last term is the Hartree contribution.
As is clear from Eq.~(\ref{eq:Ehf}), when $N_0 V_0 > 1$, it
is energetically favorable for both layers to acquire a
spontaneous spin polarization $|m_1| = |m_2| = 1$, while for
$N_0 V_0 < 1$, the layer remain unpolarized.
This result is simply the Stoner criterion for 2D itinerant
magnetism.\cite{Doniach}

\begin{figure}
\psfig{file=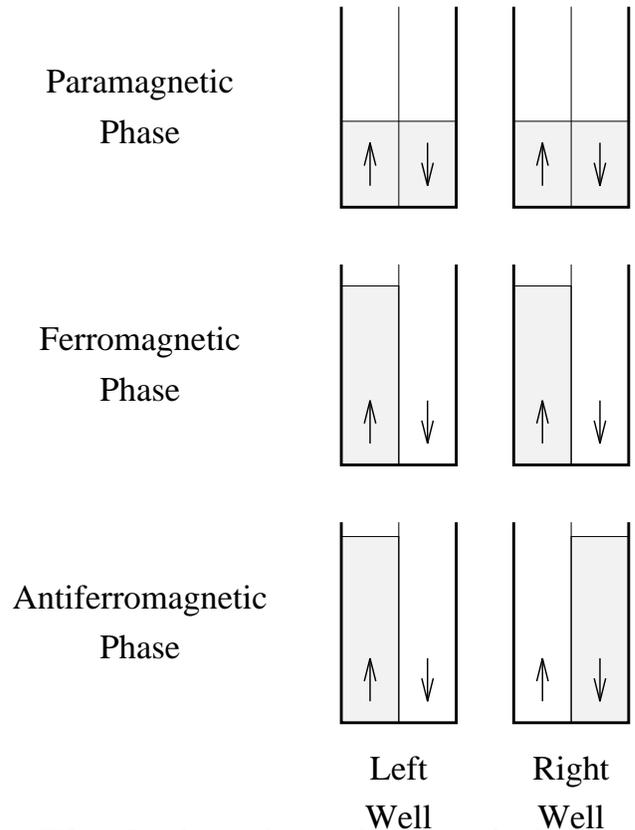,width=0.95\linewidth}
\caption{Population of spin and well states for the phases of
a double quantum well considered in Sec.~\protect\ref{sec:origin}.
In the paramagnetic phase, there is no spin polarization in either
well.
In the magnetic phases, each well is completely spin-polarized
with the polarization vectors parallel in the ferromagnetic phase
and antiparallel in the antiferromagnetic phase.
\label{fig:phases}}
\end{figure}

This description of spontaneous
spin-polarization in two-dimensional electron systems 
is unrealistic both in the use of a point-contact interaction
and in the use of the Hartree-Fock approximation.
Our objective in this section is to obtain a qualitative 
understanding of the influence of 
weak electronic tunneling between the layers on 
a double-layer system when isolated single
layer systems are close to their ferromagnetic instabilities.
We postpone a realistic discussion of 
the system parameters for which the physics we address 
in this paper is likely to realized to Sec.~\ref{sec:conclusion}.
For a sufficiently strong intra-layer repulsion, then, the
exchange interaction forces both layers to spin polarize,
but the relative orientation of the polarizations is unknown.
For simplicity, let us restrict our attention to two
possibilities for the relative orientation:
parallel (ferromagnetic) alignment or antiparallel
(antiferromagnetic) alignment.
With the spin unpolarized (paramagnetic) phase, the three
possible phases for the two-layer system are shown in
Fig.~\ref{fig:phases}.

In the absence of any inter-layer coupling, the
ferromagnetic and antiferromagnetic phases are degenerate,
but this situation changes if we introduce a small amplitude
for hopping between the layers.
Writing $c_{i{\bf k}\sigma}$ for the annihilation operator
of an electron in layer $i=1,2$ with two-dimensional wave vector
${\bf k}$ and spin projection $\sigma$, this inter-layer
hopping is governed by the Hamiltonian
\begin{equation}
H_{\bot} =
  -\frac{\Delta_0}{2} \, \sum_{{\bf k}\sigma} \left[
  c^{\dag}_{1{\bf k}\sigma} c^{~}_{2{\bf k}\sigma}
  + {\rm h.c.} \right]
\label{eq:Hbot}
\end{equation}
where $\Delta_0$ is the splitting between the symmetric and 
antisymmetric single-particle eigenstates.
The leading order change in the ground state energy due to
$H_{\bot}$ can be calculated using linear response theory, 
and is proportional to the transverse pseudospin susceptibility
of double-layer systems defined in Ref.~\onlinecite{leszek}. 
For the present model with no inter-layer interactions and contact
intra-layer interactions, the time-dependent Hartree-Fock
approximation for the susceptibility gives 
\begin{equation}
\frac{\delta E }{A} = \left\{
  \begin{array}{ll}
  -\frac{1}{2} N_0 \Delta_0^2 & {\rm (paramagnetic)} \\
  -\frac{1}{4} N_0 \Delta_0^2 & {\rm (ferromagnetic)} \\
  -\frac{1}{2} \Delta_0^2/V_0 & {\rm (antiferromagnetic)}
  \end{array} \right.
\label{eq:Ebot}
\end{equation}
The fact that the results for paramagnetic and ferromagnetic 
states are independent of the interaction strength is 
a special property of the present model related to both the 
absence of interlayer interactions and the wave vector 
independence of the exchange self-energy.  

We see that, in all cases, introducing inter-layer hopping
reduces the ground state energy,
as one might expect when the confinement of the electrons
to the layers is weakened.
Comparing the energies of the ferromagnetic and
antiferromagnetic phases, we also observe that the
degeneracy between these phases is broken by the hopping term.
Specifically, the antiferromagnetic phase is found to be more
stable than the ferromagnetic phase if $N_0 V_0 < 2$,
implying that the inter-layer hopping opens up a region of
antiferromagnetic order between the paramagnetic and
ferromagnetic phases.
The mechanism for this stabilization can be deduced from the
form of $\delta E  / A$ to be a superexchange
interaction; that is, an electron is able to hop from one
layer to the other and back at the cost of a Hubbard energy
in the intermediate state, yielding an energy savings of
$\Delta_0^2 / 2 V_0$.
A similar mechanism is blocked by the Pauli exclusion principle 
in the ferromagnetic phase , since the hopping
Hamiltonian preserves spin [Eq.~(\ref{eq:Hbot})].

We argue that this mechanism favoring an antiferromagnetic arrangement
of the ordered moments in the two layers will be dominant 
in most circumstances.  
The calculations described below indicate that an effective
superexchange interaction between the wells should
stabilize an antiferromagnetic phase for moderate
interaction strengths.
This explanation of the zero-field antiferromagnetic phase
suggests why the presence of asymmetry and additional
interaction matrix elements may not eliminate this phase:
both the two-dimensional ferromagnetism within each well
and the superexchange interaction between the wells should
be fairly insensitive to these perturbations.
In the remainder of the paper, we perform a more
detailed self-consistent Hartree-Fock calculation to support
this statement and to explore the consequences of these
perturbations.

\section{FORMALISM}
\label{sec:formalism}

To accomplish the goal of investigating matrix element
and asymmetry effects, we employ an extension of the
point-contact model described in Ref.~\onlinecite{rad-das}.
In the original calculation, the full Coulomb interaction
between the electrons in the double quantum well was
approximated by a delta function in real space with only the
matrix elements of this interaction between the lowest two
subbands being kept and the remainder being set to zero.\cite{rad-das}
The use of a delta-function or point-contact interaction is
based primarily on a desire to create a simple, solvable
model which mimics the qualitative features of the fully
interacting system.  A quantitative theory would require 
that the interaction be made realistic, and also that the 
interactions be treated more accurately than in the 
Hartree-Fock approximation.  In practise this would require 
quantum Monte-Carlo calculations of some type, which would involve
an enormous amount of effort and would not be able to address 
the excitation spectrum which provides the experimental signature
for the state we are proposing.  
We therefore maintain the point-contact form of the interaction here.
However, there is no reason to set matrix elements
of this interaction other than those between the lowest subbands
to zero, as in previous work.
Indeed, we shall see below that these other matrix elements
are of the same order as the inter-subband ones.
Thus, we shall include all the matrix elements between the
lowest two subbands in our calculations.

\begin{figure}
\psfig{file=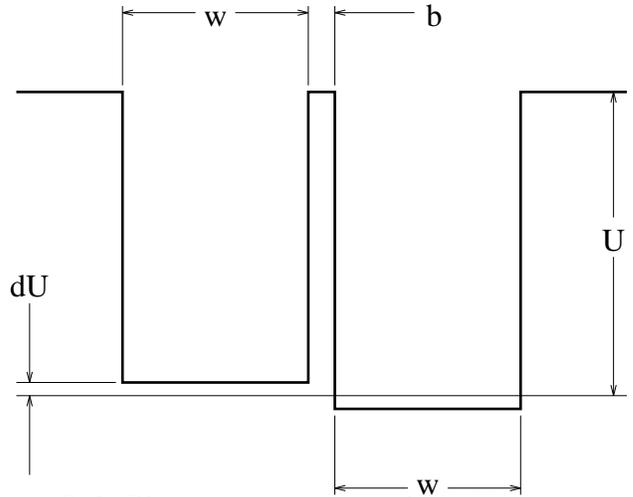,width=0.95\linewidth}
\caption{Diagram of the asymmetric double quantum well
considered in this paper indicating the well width $w$, barrier
width $b$, confining potential $U$, and potential asymmetry
$dU$.
The values of the parameters used here and in previous
studies of magnetic instabilities of these
systems\protect\cite{das-tam,rad-das} model a typical
GaAs/Al$_x$Ga$_{1-x}$As double quantum well with
$w$ = 140~\AA, $b$ = 30~\AA, and $U$ = 220~meV.
We consider both symmetric ($dU = 0$) and weakly
asymmetric ($dU = 0.5$~meV) double quantum wells in what
follows.
\label{fig:dqw}}
\end{figure}

In addition to the issue of intra-subband repulsion, we would
also like to study the effects of quantum well asymmetry on
the phase diagram and collective modes of this system.
This asymmetry arises in real quantum wells through alloy
fluctuations across the profile of the well or fluctuations
in the well thickness, and we model it by allowing one of
the wells to be deeper than the other, as illustrated in
Fig.~\ref{fig:dqw}.
Although we have assumed the effective asymmetry to enter
through the well depth, the results of our calculations for
systems with asymmetric well widths should be qualitatively
similar.

We therefore consider a three-dimensional electron gas confined along
the $z$-direction by a potential $V_{CONF}(z)$ of the type
shown in Fig.~\ref{fig:dqw} which interacts through
a {\it three dimensional} 
point-contact potential $V({\bf R}) = V \delta({\bf R})$.
In the absence of the interaction $V({\bf R})$, the
electronic eigenstates are given by the solution of the
time-independent Schr\"{o}dinger equation
\begin{equation}
\left[ -\frac{\hbar^2}{2m^*} \frac{d^2}{dz^2} + V_{CONF} (z) \right]
  \xi_n (z) = \epsilon_n \xi_n (z),
\label{eq:xidef}
\end{equation}
where $m^*$ is the effective mass, which is assumed to be 
constant throughout the heterostructure.
Normalizing these eigenfunctions by
\begin{equation}
  \int dz \left| \xi_n(z) \right|^2 = 1,
\label{eq:norm}
\end{equation}
we write the electron annihilation operator
$\psi_{\sigma} ({\bf R})$ as
\begin{equation}
\psi_{\sigma} ({\bf R}) =
  \frac{1}{\sqrt{A}} \sum_{n{\bf k}} e^{i{\bf k \cdot r}} \xi_n (z)
  c_{n{\bf k}\sigma},
\label{eq:basis}
\end{equation}
where ${\bf R} = ({\bf r},z) = (x,y,z)$, ${\bf k} = (k_x, k_y)$,
$A$ is the transverse area of the sample, and $c_{n{\bf k}\sigma}$
annihilates a quasiparticle in subband $n$, of transverse wave vector
${\bf k}$, and with spin projection $\sigma$ (these conventions will
be used throughout this paper).

Defining a composite subband and spin index $a = (n_a, \sigma_a)$
with summation over repeated indices implied, the Hamiltonian
for this system is
\begin{eqnarray}
H &=& \sum_{\bf k} \, \epsilon_{a{\bf k}}^{~} \,
  c^{\dag}_{a {\bf k}} c^{~}_{a {\bf k}} \nonumber \\
&&+ \frac{1}{2 A} \, \sum_{{\bf k} {\bf k'} {\bf q}}
  \, V_{ad,bc} \,
  c^{\dag}_{a {\bf k + q}} c^{\dag}_{b {\bf k' - q}}
  c^{ }_{c {\bf k'}} c^{ }_{d {\bf k}} .
\label{eq:H}
\end{eqnarray}
Here, the quasiparticle energy
\begin{equation}
\epsilon_{a{\bf k}} = \epsilon_n + \frac{\hbar^2 k^2}{2m^*} - \mu
\label{eq:ea}
\end{equation}
is measured with respect to the chemical potential $\mu$
and the matrix elements of the interaction are
\begin{eqnarray}
V_{ab,cd} &=&
  \delta_{\sigma_a \sigma_b} \delta_{\sigma_c \sigma_d} \,
  V \, \int dz \,\xi_{n_a}^{*} (z) \xi_{n_b}^{~} (z) \,
 \xi_{n_c}^{*} (z) \xi_{n_d}^{~} (z) .
\label{eq:V}
\end{eqnarray}

We treat this Hamiltonian within self-consistent
Hartree-Fock theory allowing for the possibility of
phases with broken symmetry in subband and spin indices, 
but imposing translational invariance within each layer.
The electronic Green's function in the interacting system can
therefore be written
\begin{equation}
G_{ab} (k_n) = - \int_0^{\beta} d\tau \, e^{i\omega_n \tau} \,
  \left< T_\tau \left[ c_{a{\bf k}} (\tau) c_{b{\bf k}}^{\dag} (0)
  \right] \right>,
\label{eq:G}
\end{equation}
where $k_n = ({\bf k},i\omega_n)$, $\beta = 1 / T$
($\hbar = k_B = 1$ throughout this paper), and the rest of the
notation is standard.\cite{Mahan}
This Green's function is determined self-consistently from
the self-energy in the Hartree-Fock approximation,
\begin{equation}
\Sigma_{ab} =
  \left[ V_{ab,dc} - V_{ac,db} \right] \,
  \frac{T}{A} \sum_{k^{\prime}_m} e^{-i \omega_m 0-} \,
  G_{cd} (k^{\prime}_m),
\label{eq:Sigma}
\end{equation}
and the Dyson equation,
\begin{equation}
\left[ \left( i\omega_n - \epsilon_{a{\bf k}} \right) \delta_{ab}
  - \Sigma_{ab} \right] \, G_{bc} (k_n) = \delta_{ac},
\label{eq:Dyson}
\end{equation}
under the constraint of constant sheet density $N_s$,
\begin{equation}
N_s = \frac{T}{A} \sum_{k_m} e^{-i \omega_m 0-} \, G_{aa} (k_m) ,
\label{eq:filling}
\end{equation}
which determines the chemical potential.

We solve these equations in the following way.
Given a double quantum well structure defined by the
effective mass $m^*$, sheet density $N_s$, and the
structural parameters shown in Fig.~\ref{fig:dqw}, we
compute the eigenfunctions and eigenenergies by solving
Eq.~(\ref{eq:xidef}) with the normalization condition
Eq.~(\ref{eq:norm}).
These eigenfunctions are used to compute the matrix elements
of the interaction via Eq.~(\ref{eq:V}) in terms of a single
parameter V.
The resulting matrix elements and eigenenergies are employed
to solve Eqs.~(\ref{eq:G})-(\ref{eq:filling}) self-consistently
at $T = 0$ following the procedure outlined in Sec.~II.B of
Ref.~\onlinecite{rad-das} and including {\it all} matrix
elements of the interaction between the lowest two subbands.
This procedure yields the ground state properties of the
system as a function of the sheet density $N_s$, the
interaction parameter $V$, and the structural parameters of
the double quantum well [Fig.~\ref{fig:dqw}].

To illuminate the properties of the interacting system further,
we also compute the generalized density-density response function,
which is defined by the analytic continuation of
\begin{eqnarray}
\Pi^{\mu} ({\bf Q},i\nu_n) &=&
 - \int \frac{d{\bf R}}{\cal V} \, e^{-i{\bf Q \cdot R}} \,
  \int_0^{\beta} d\tau \, e^{i\nu_n\tau} \times \nonumber \\
&&  \left< T_{\tau} \left[ \rho^{\mu}({\bf R},\tau) \,
  \rho^{\mu}({\bf 0},0) \right] \right>
\label{eq:Pi}
\end{eqnarray}
to real frequencies.
In this expression, ${\cal V}$ is the system volume and the
generalized density operator is
\begin{equation}
\rho^{\mu}({\bf R}) =
  \sum_{\sigma\sigma'} \, \psi^{\dag}_{\sigma}({\bf R})
  \sigma^{\mu}_{\sigma\sigma'} \psi^{~}_{\sigma'}({\bf R})
\label{eq:dop}
\end{equation}
with $ \psi_{\sigma}({\bf R})$ given by
Eq.~(\ref{eq:basis}) and
$\sigma^{\mu} = (1,\sigma^x,\sigma^y,\sigma^z)$ are the
Pauli matrices.
This response function is computed from the non-interacting
response function in subband and spin space within the
conserving approximation described in
Sec.~V.A of Ref.~\onlinecite{rad-das} but with the inclusion
of all interaction matrix elements.
In addition to the information this response function
reveals about the excitations of the interacting system,
its imaginary part is proportional to the intensity observed
in resonant inelastic light scattering
measurements,\cite{pin-abs,das:ils} allowing us to make
contact with experiment.
This is particularly relevant here, because searches for
the antiferromagnetic phase in both finite\cite{Pinczuk} and
zero\cite{Plaut} magnetic field have employed this
technique.

\section{RESULTS}
\label{sec:results}

In this Section, we apply the formalism described in
Sec.~\ref{sec:formalism} to compute the ground state phase
diagram and collective modes in a typical GaAs/Al$_x$Ga$_{1-x}$As
double quantum well structure which is expected to exhibit
the zero-field antiferromagnetic instability.
The structure has a well width of 140~\AA, a barrier width
of 30~\AA, and a well depth of 220~meV [cf.
Fig.~\ref{fig:dqw}], and an electronic effective mass
$m^* = 0.067\,m_e$.
For the moment, we leave the asymmetry unspecified.

\subsection{Matrix Elements}
\label{sec:me}

As a first step in obtaining the phase diagram for this
structure, we must solve the time-independent
Schr\"{o}dinger equaution [Eq.~(\ref{eq:xidef})] for the
lowest two eigenfunctions $\xi_n (z)$ and eigenenergies
$\epsilon_n$ at a fixed value of the asymmetry parameter
$dU$ [cf. Fig.~\ref{fig:dqw}] and then compute the matrix
elements of the interaction through Eq.~(\ref{eq:V}).
Solving Eq.~(\ref{eq:xidef}) is straightforward and yields a
splitting of the non-interacting eigenstates of
$\Delta^0_{\rm SAS} = \epsilon_2 - \epsilon_1 = 2.25$~meV
for $dU = 0$.
As $dU$ is increased, this splitting increases to a
maximum of 18.5~meV at $dU = 9.4$~meV; for larger $dU$, the
lowest eigenfunction is localized in one well.
Since the structures examined experimentally have subband
splittings on the order of 1~meV,\cite{Pinczuk,Plaut}
 we restrict our attention to small $dU$ values.

\begin{figure}
\psfig{file=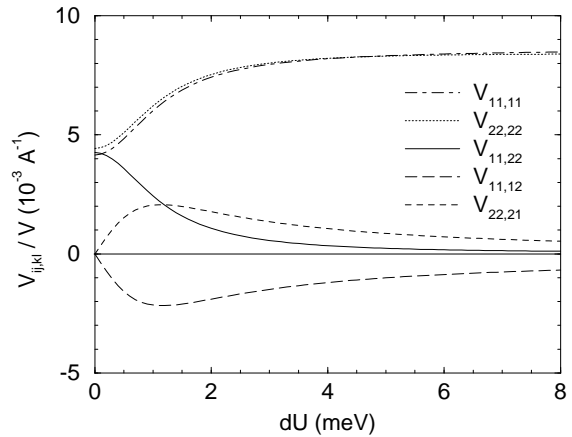,width=0.95\linewidth}
\caption{Dependence of the interaction matrix elements
$V_{ij,kl} / V$ [Eq.~(\protect\ref{eq:V})] on the double
quantum well asymmetry parameter $dU$ [cf.
Fig.~\protect\ref{fig:dqw}].
Shown are the intra-subband matrix elements $V_{11,11}$
(dot-dashed line) and $V_{22,22}$ (dotted line), the
inter-subband matrix element $V_{11,22}$ (solid line),
and the asymmetry-induced matrix elements $V_{11,12}$
(long dashed line) and $V_{22,21}$ (short-dashed line).
For the point-contact interaction employed in this paper,
the matrix elements are invariant under permutation of the
indices, so these matrix elements span the entire set.
\label{fig:me}}
\end{figure}

The dependence of the matrix elements on the asymmetry
parameter is somewhat more interesting and merits a brief
discussion.
Since our model interaction is a delta function in real
space and we have chosen the wave functions to be real,
the matrix elements defined by Eq.~(\ref{eq:V}) are invariant
under permutation of the indices.
Thus, there are only five independent matrix elements,
$V_{11,11}$, $V_{22,22}$, $V_{11,22}$, $V_{11,12}$, and
$V_{22,21}$, which are displayed in Fig.~\ref{fig:me} as a
function of the asymmetry parameter $dU$.
In a symmetric double quantum well ($dU = 0$), the
$V_{11,12}$ and $V_{22,21}$ matrix elements vanish by
symmetry, but the remaining inter-subband ($V_{11,22}$) and
intra-subband ($V_{11,11}$ and $V_{22,22}$) matrix elements
are equal to within 5~\%.  We remark that for more 
realistic interaction models $V_{11,22}$, which is 
roughly\cite{boebinger} proportional to the difference of 
intra-layer and inter-layer interactions, is weaker 
than $V_{11,11}$ and $V_{22,22}$ which are roughly 
proportional to the sum.
The latter matrix elements would be equal if the
electrons were localized to the wells; the fact that they
are nearly so indicates that the wave function overlap
between the wells is small.
Additionally, this calculation provides direct evidence that
the neglect of the intra-subband repulsion employed in
earlier work\cite{rad-das} is not generally 
justified for these double quantum well structures.
We shall see, however, that their inclusion in the
calculation changes the qualitative picture only slightly.

As the asymmetry is increased from zero, we discern several
features.
We see that the intra-subband matrix elements $V_{11,11}$ and
$V_{22,22}$ are approximately equal and increase with increasing
$dU$ to saturate at a value about twice the $dU = 0$ one.
The approximate equality of these diagonal matrix elements
follows from the normalization condition, Eq.~(\ref{eq:norm}),
imposed on the two eigenfunctions.
The increase in these matrix elements with $dU$, on the
other hand, can be attributed to the increasing confinement
of the wave functions of the two subbands to opposite wells,
similar to what occurs in the presence of an applied electric
field.\cite{Katayama,Ying}
Thus, at large $dU$, the two subband wave functions are almost
completely localized in opposite wells, enhancing the
magnitude of the diagonal matrix elements.
For the same reason, the inter-subband matrix element
$V_{11,22}$ decreases with increasing asymmetry:
as the wave functions from different subbands are
increasingly localized, their overlap, and hence $V_{11,22}$,
decreases to zero.
Since the zero-field antiferromagnetic transition depends on
this matrix element, it is clear that large asymmetry is
inimical to this phase.
Based on the small values of the observed
splitting of the lowest two subbands in the experimental
samples,\cite{Pinczuk,Plaut} however, we expect the actual
samples to be in a regime in which $V_{11,22}$ is still non-negligible.
Hence, the antiferromagnetic phase is not immediately excluded.

Finally, we note that the mixing terms $V_{11,12}$ and
$V_{22,21}$ have opposite signs and increase in magnitude
with $dU$ to a maximum around $dU = 1$~meV.
The wave function for the lowest ($n = 1$) subband has 
no nodes and we chose it to be 
positive. Orthogonality requires that the 
$n = 2$ wave function have a node and we choose it to
be negative in the well where
$|\xi_1(z)|^2$ is largest; with this convention $V_{11,12}$ is 
negative and $V_{22,21}$ is positive, as observed in
Fig.~\ref{fig:me}.
In addition, since these matrix elements must vanish both
in the symmetric ($dU = 0$) limit and when $dU$ is large and the
subband wave functions are localized in different wells, the
maximum seen in this figure is also expected.
These results suggest that the mode coupling between intra-
and inter-subband excitations induced by these matrix
elements will be maximal around $dU = 1$~meV.

Taken together, the behavior of the interaction matrix
elements presented in Fig.~\ref{fig:me} indicate that the
zero-field antiferromagnetic phase will probably not be
stable against large asymmetry in the quantum wells.
However, the current experimental samples have subband
splittings more consistent with small asymmetry,
and therefore these samples may be of high enough quality
to observe this phase, at least in principle.
To examine this situation further, we shall compute the phase
diagram and collective modes for two choices of $dU$ in the
weak asymmetry regime:  $dU = 0$ (the symmetric case) and
$dU = 0.5$~meV (the asymmetric case).
The following subsections discuss the results of these
calculations.

\subsection{Phase Diagram}
\label{sec:phases}

\begin{figure}
\psfig{file=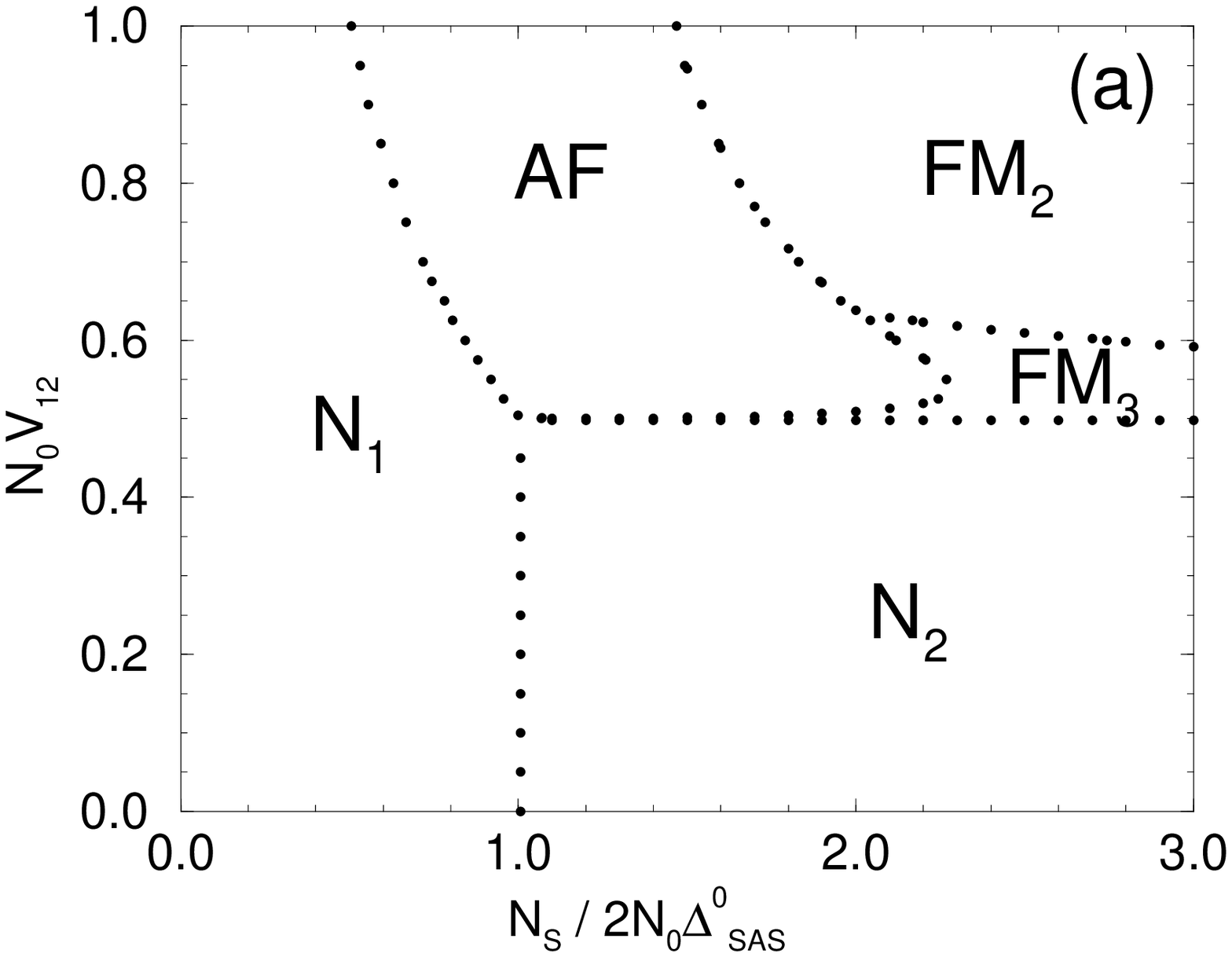,width=0.95\linewidth}
\psfig{file=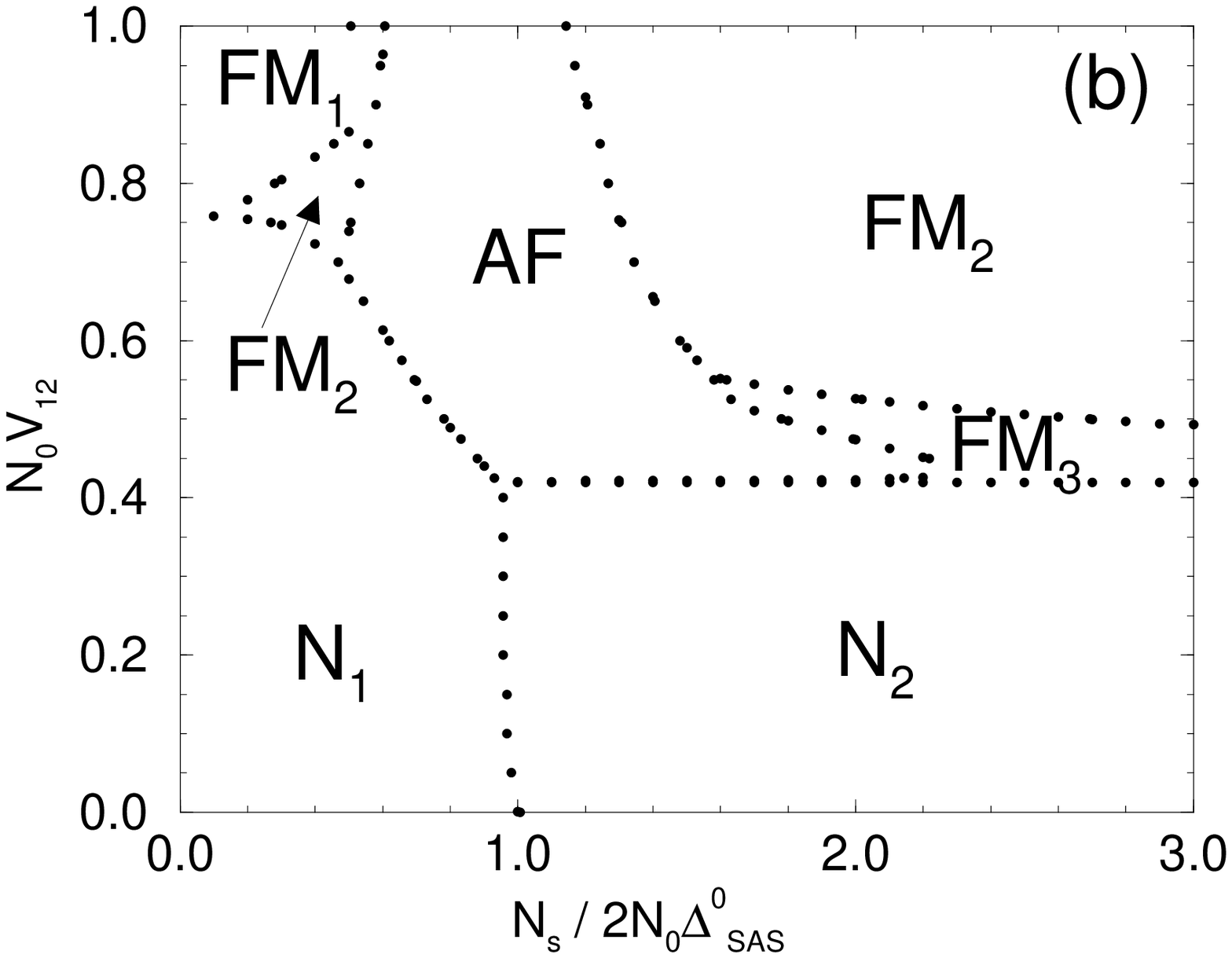,width=0.95\linewidth}
\caption{Phase diagram of (a) a symmetric and (b) an asymmetric
double quantum well in terms of the sheet density $N_s$ and
the interaction parameter $V_{12}$ for the point-contact model
described in Sec.~\protect\ref{sec:formalism}.
The points are the numerically computed boundaries
between the various phases of this model which are labeled
as follows:
$N_1$ and $N_2$ are paramagnetic phases with one and two
spin-degenerate subbands occupied, respectively;
$FM_i$ are ferromagnetic phases with $i$ non-degenerate
subbands occupied; and
$AF$ is the antiferromagnetic phase.
See Fig.~\protect\ref{fig:phases} for a real-space depiction of
these phases and Fig.~\protect\ref{fig:dqw} for the
structural parameters of the double quantum well.
In the figure, $N_0$ is the single-spin, 2D electronic
density of states, and $\Delta^0_{\rm SAS}$ is the
splitting between the lowest two states in the non-interacting
system.
Note that asymmetry reduces the extent of the antiferromagnetic
phase in (b) but does not destroy it altogether.
\label{fig:phasediagram}}
\end{figure}

As described in the preceding subsection, the structure of
the double quantum well yields the eigenenergies and
eigenfunctions of the non-interacting system which are then
used to compute the interaction matrix elements up to an
overall factor $V$ [Eq.~(\ref{eq:V})].
To be consistent with earlier work,\cite{rad-das} we choose
to parameterize the interaction strength by the magnitude of
the inter-subband repulsion $V_{11,22} = V_{12}$ rather than
by $V$, but all the matrix elements are uniquely determined
by either parameter.
With the structure and interaction strength fixed, the only
other parameter in our model is the sheet density $N_s$.
Given these parameters, Eqs.~(\ref{eq:G})-(\ref{eq:filling})
can be solved at $T = 0$ to yield the interacting ground state
of the system.
The resulting phase diagrams in terms of the dimensionless
interaction strength $N_0 V_{12}$ and sheet density $N_s / 2
N_0 \Delta^0_{\rm SAS}$ are presented in
Fig.~\ref{fig:phasediagram} for the double quantum well
structure of Fig.~\ref{fig:dqw}.
In these figures, $N_0$ is the single-spin, two-dimensional
density of states, and $\Delta^0_{\rm SAS} = \epsilon_2 -
\epsilon_1$ is the subband splitting in the non-interacting
system.

Consider the symmetric ($dU = 0$) case first
[Fig.~\ref{fig:phasediagram}(a)].
At first glance, this phase diagram is very similar to the
one obtained from the neglect of intra-subband
repulsion,\cite{rad-das} even though this repulsion is
included in our calculations.
At low density, the paramagnetic phase with one
spin-degenerate subband occupied (labeled $N_1$ in the
figure) is stable, while at higher densities ($N_s /
2N_0\Delta^0_{\rm SAS} > 1$) but weak interaction ($N_0
V_{12} < 1/2$) the paramagnetic phase with two
spin-degenerate subbands occupied (labeled $N_2$) is stable.
For larger interaction strengths and intermediate densities,
we see a broad region in which the antiferromagnetic ($AF$)
phase obtains.
Thus, the inclusion of intra-subband repulsion
does not eliminate the zero-field antiferromagnetic phase
from the phase diagram of the model, despite the fact that
the intra- and inter-subband repulsions are of the same
order [Fig.~\ref{fig:me}].

These matrix elements do have an effect on the phase
diagram, however.
At higher densities ($N_s/2N_0\Delta^0_{\rm SAS}$ larger
than approximately 2) and strong interaction ($N_0V_{12} >
1/2$), two different ferromagnetic phases appear.
In these phases, each quantum well is spin polarized, and
the polarization vectors are parallel.
They differ in the arrangement and filling of the
interacting bands, as indicated by the labels in
the figure:  the phase $FM_i$ corresponds to an interacting
band structure with $i$ spin-split subbands occupied.
The presence of the ferromagnetic phases is expected when
intra-subband repulsion in present; what is surprising is
that the ferromagnetic phases do not exclude the presence of
the antiferromagnetic phase.
As argued in Sec.~\ref{sec:origin} based on a weak-coupling
model, the antiferromagnetic phase is stabilized by a
superexchange interaction for intermediate interaction
strengths.
Similar behavior is seen in Fig.~\ref{fig:phasediagram}(a)
based on our strong-coupling computations and presumably
originates from the same mechanism.
Note that, in the limit of vanishing hopping between the
wells, $V_{12}$ may be associated with $V_0/2$ in the model
of Sec.~\ref{sec:origin}, implying that the phase boundary
$N_0 V_{12} = 1/2$ in Fig.~\ref{fig:phasediagram}(a) is nothing
but the Stoner criterion for the related Hubbard model.
The fact that the model of Sec.~\ref{sec:origin} does not
have a closer correspondence to
Fig.~\ref{fig:phasediagram}(a) suggests that the
relative magnitudes of the intra- and inter-subband
interaction matrix elements, which are all equal in
the model of Sec.~\ref{sec:origin}, are important for
determining whether ferro- or antiferromagnetic phases
obtain in a specific region of the phase diagram.

With the introduction of weak asymmetry ($dU = 0.5$~meV),
the qualitative features of the phase diagram do not change,
as seen in Fig.~\ref{fig:phasediagram}(b).
As before, we find paramagnetic phases at small interaction
strengths, the antiferromagnetic phase at larger interaction
strengths and intermediate densities, and ferromagnetic
phases at larger interaction strengths and higher densities.
The identification of these phases precisely matches those
in the symmetric case of Fig.~\ref{fig:phasediagram}(a),
although the position of the phase boundaries have shifted
somewhat.
An apparently new feature occurs at low density and large
interaction strength, where an $FM_1$ ferromagnetic phase has
replaced the paramagnetic $N_1$ phase.
However, this phase also occurs in the symmetric structure
when $N_0 V_{12}>1$, but is cut-off in
Fig.~\ref{fig:phasediagram}(a).
It corresponds to the usual ferromagnetic instability in a
single, spin-degenerate band which obtains when the
interaction is sufficiently strong.\cite{Doniach}

Although weak asymmetry clearly does not destroy the
antiferromagnetic phase, it does have observable
consequences.
The most noticeable effect of the asymmetry on the ground
state is that the spin polarizations in the magnetic phases,
which can be obtained from the expectation value of the
density operator, Eq.~(\ref{eq:dop}), are no longer of equal
magnitude in the wells.
This is an obvious consequence of an asymmetric structure
which nonetheless does not disturb the identification of
ferro- and antiferromagnetic phases, since one can determine
whether the spin polarizations are parallel or antiparallel
without referring to their magnitudes.

\subsection{Collective Modes}
\label{sec:modes}

The effects of asymmetry on the antiferromagnetic phase
cannot be fully appreciated based on the ground-state
properties alone, but must be augmented by an examination of
the excitation spectrum of the system.
We focus on the spin-density excitations in what follows,
since they are the excitations most strongly coupled to the
magnetic nature of the ground state and can also be probed
experimentally by Raman scattering.\cite{pin-abs,das:ils}
For these calculations, we compute the spin-spin
response function ($\mu = 3$ in Eq.~(\ref{eq:Pi}))
for the appropriate ground state as discussed in
Sec.~\ref{sec:formalism} and identify the collective modes
by peaks in the imaginary part of this response function.
Since this procedure is used in the Raman scattering
measurements, our results have direct implications
for experiment and we shall discuss them in this context.
For concreteness, we fix the interaction strength and sweep
the sheet density through the second-order transition from
the $N_1$ to the $AF$ phase in these calculations [cf.
Fig.~\ref{fig:phasediagram}].

\begin{figure}
\psfig{file=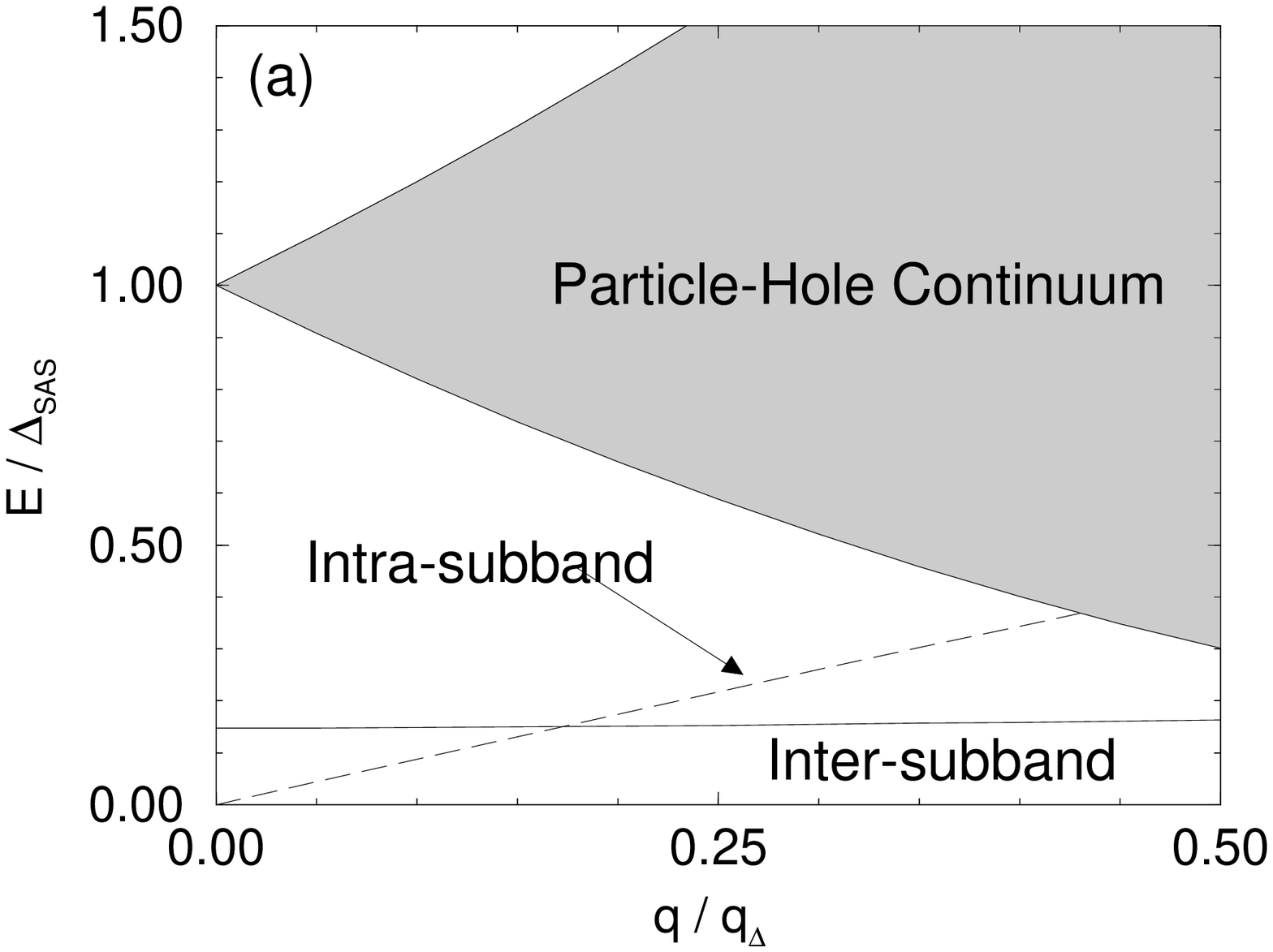,width=0.95\linewidth}
\psfig{file=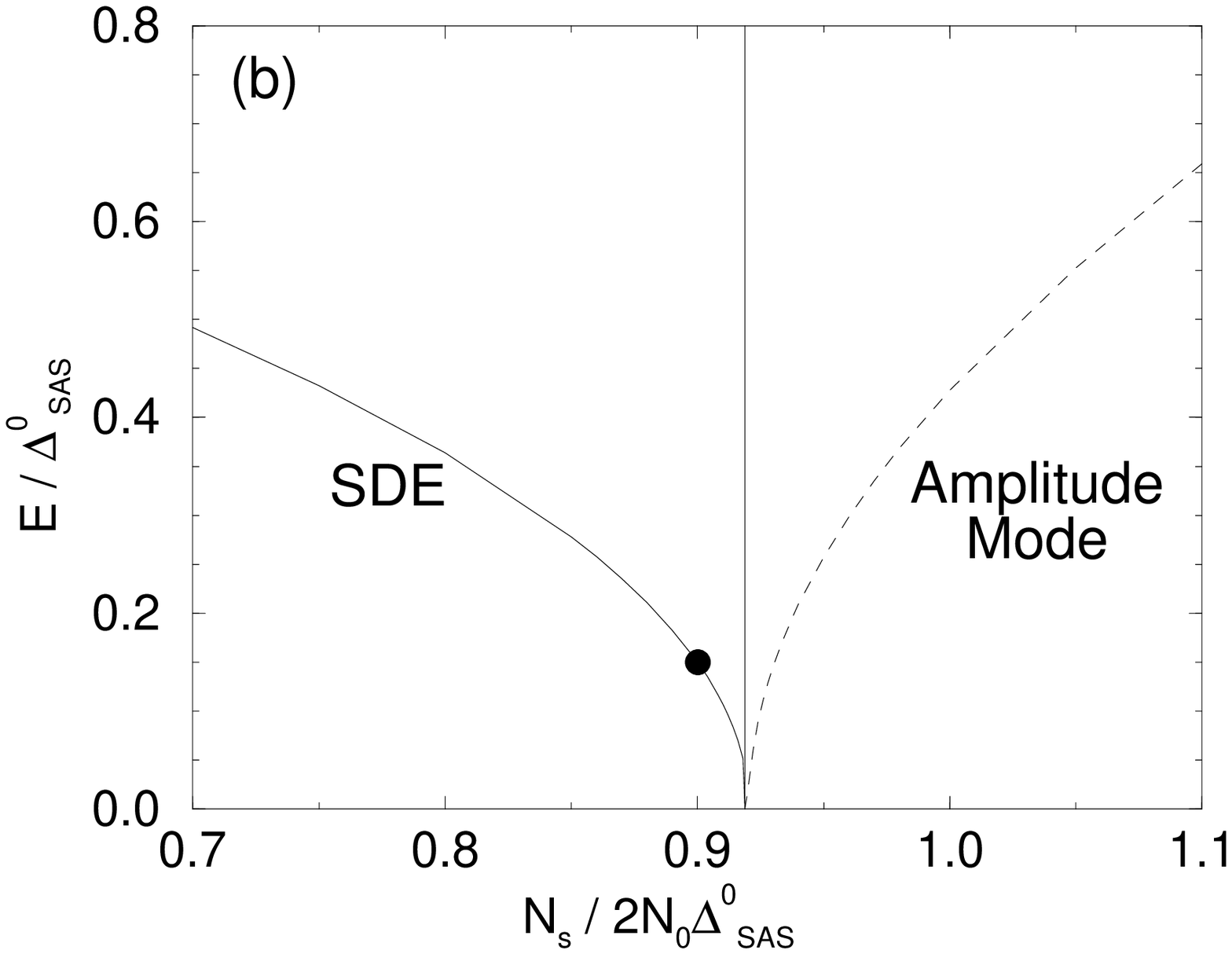,width=0.95\linewidth}
\caption{Collective spin-density excitations in a symmetric
double quantum well of the type shown in Fig.~\protect\ref{fig:dqw}
with $dU = 0$.
(a) Collective mode spectrum in terms of excitation energy
$E$ and intra-well wave vector $q$ showing the
inter- (solid line) and intra- (dashed line) collective spin-density
excitations as well as the continuum of inter-subband particle-hole
excitations (shaded area).
Although not apparent in the figure, the inter-subband
collective mode does disperse with $q$.
(b) $q = 0$ inter-subband spin density excitation (SDE) energy
$E$ as the sheet density $N_s$ is tuned through the $N_1$ to $AF$
transition at fixed interaction strength $N_0 V_{12} = 0.55$
[cf. Fig.~\protect\ref{fig:phasediagram}(a)].
Note that the SDE (solid line) softens completely at the
transition point and that the collective mode appearing on the
antiferromagnetic side ($N_s / 2N_0\Delta^0_{\rm SAS} >
0.919$) is the amplitude mode of the antiferromagnetic order
parameter (dashed line).
The dot indicates the point in parameter space presented in
(a).
In this figure, $N_0$ is the single-spin, 2D electronic
density of states, $\Delta^0_{\rm SAS}$ is the
energy separation of the lowest two subbands
in a non-interacting double quantum well, $\Delta_{\rm SAS}$
is the splitting in the interacting system, and
$q_\Delta^2 = 2 m^* \Delta^0_{\rm SAS} / \hbar^2$.  The 
intra-band particle-hole continuum is not indicated.
\label{fig:symmode}}
\end{figure}

As an introduction to the general phenomenology of
spin-density excitations in double quantum wells, consider
the symmetric ($dU = 0$) case first.
By appropriately arranging the light scattering geometry,
Raman scattering can selectively probe the intersubband
spin-density excitations,\cite{Pinczuk,Plaut} which, in our
approximation, have the
form shown in Fig.~\ref{fig:symmode}(a).
In addition to a continuum of inter-subband particle-hole
excitations, there is a collective spin-density excitation
(SDE) with a finite energy at $q = 0$ which disperses with
increasing $q$ toward the particle-hole continuum.
The magnitude of the $q = 0$ SDE energy is reduced from the
subband splitting $\Delta_{\rm SAS}$ by vertex corrections
appearing in the response function due the the exchange
interaction.\cite{Ando-RMP,pinczuk89}
In addition, there is an intra-subband SDE which has a
linear-in-q dispersion in our model, shown by the dashed
line in Fig.~\ref{fig:symmode}(a).
If the well is symmetric, this mode will not appear in
Raman spectra taken in a scattering geometry meant to
observe inter-subband excitations.
A symmetric system with identical quantum wells is 
invariant under inversion about the mid-point between
the wells so that all 
states can be classified by a parity quantum number.
Inter-subband excitations, which are odd, and intrasubband 
excitations, which are even, do not interact and can cross as seen
in Fig.~\ref{fig:symmode}(a).

As the density increases in our model, the exchange-induced
reduction in the $q=0$ inter-subband SDE energy increases
until the mode softens entirely, as illustrated in
Fig.~\ref{fig:symmode}(b).
This complete softening was seen initially in time-dependent,
local density approximation calculations of the SDE spectrum in
these systems and was the first evidence for the zero-field
antiferromagnetic phase.\cite{das-tam}
An analysis of the real-space spin response identified this
softening as an antiferromagnetic transition of the
well spin polarizations.\cite{rad-das}
As the density is increased past the antiferromagnetic
transition, the inter-subband SDE turns into the collective
mode associated with amplitude fluctuations of the
antiferromagnetic order parameter.\cite{rad-das}
Experimentally, then, one expects to see a complete
softening of the inter-subband SDE and the recovery of
this amplitude mode as the density is tuned through
the transition.

\begin{figure}
\psfig{file=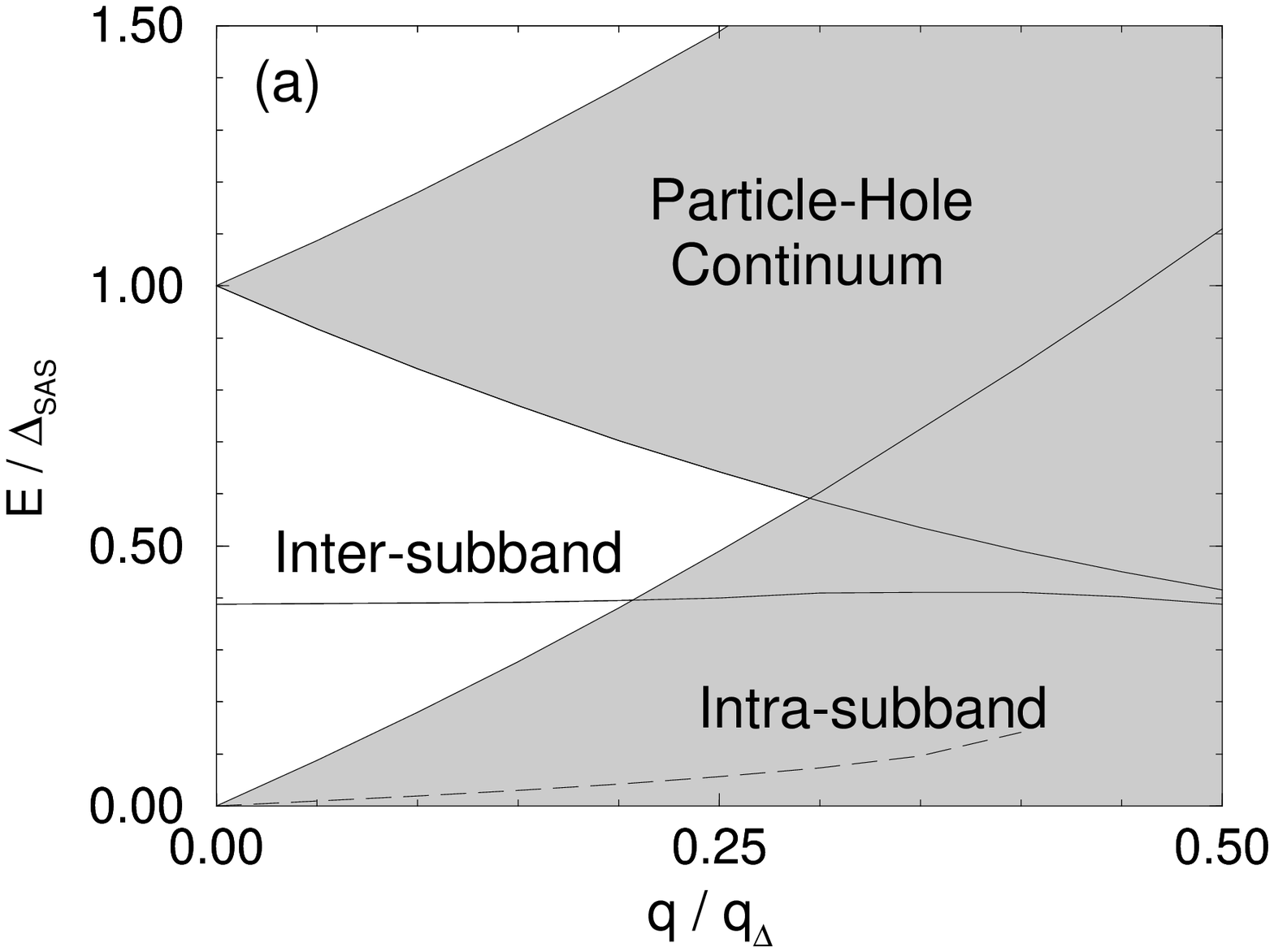,width=0.95\linewidth}
\psfig{file=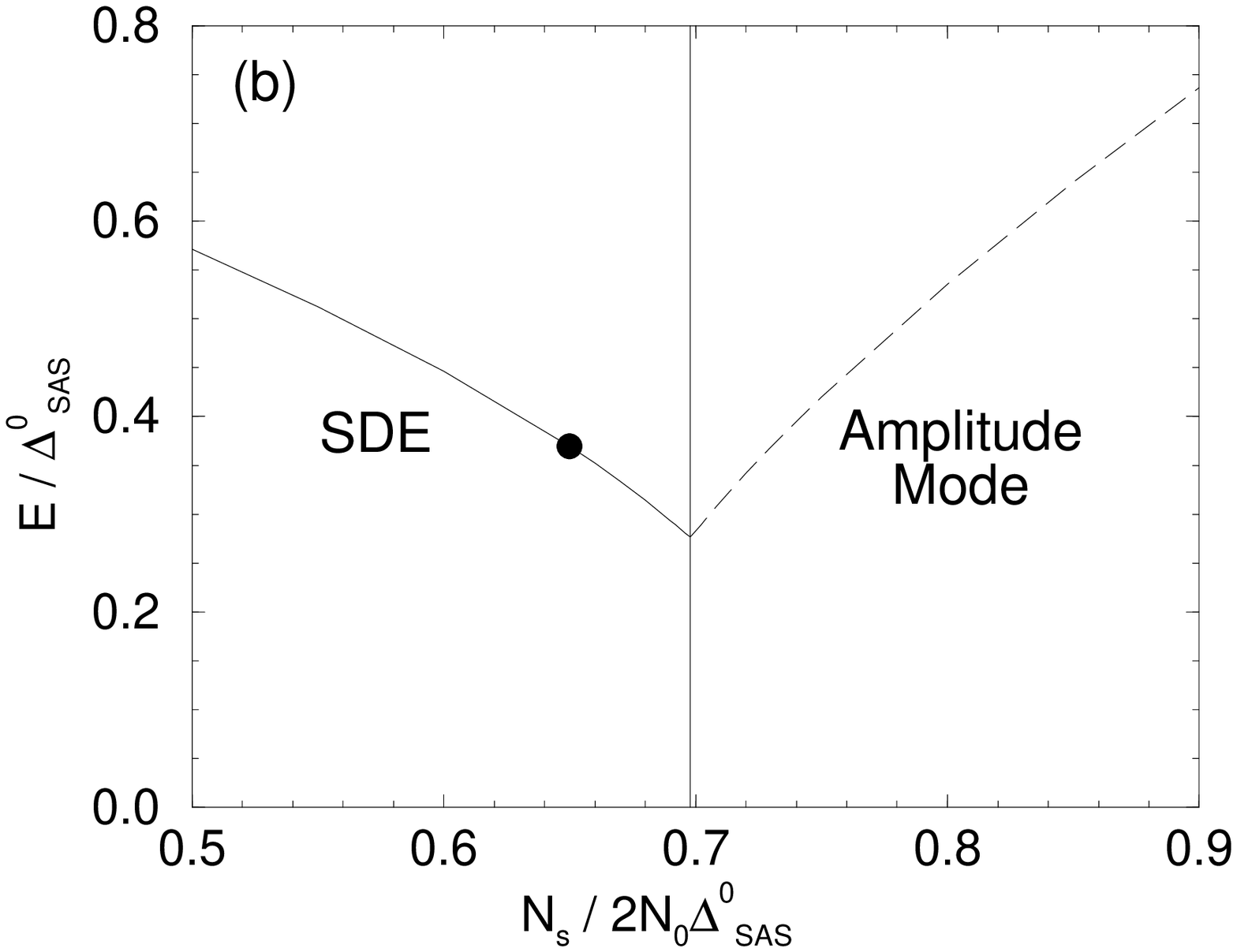,width=0.95\linewidth}
\caption{Collective spin-density excitations in a weakly
asymmetric double quantum well of the type shown in
Fig.~\protect\ref{fig:dqw} with $dU = 0.5$~meV.
The organization of and notation in the figure are the same
as Fig.~\protect\ref{fig:symmode} with the interaction strength
fixed at $N_0 V_{12} = 0.55$ and the antiferromagnetic
transition occurring at $N_s / 2N_0\Delta^0_{\rm SAS} = 0.698$.
As seen in (a), the asymmetry induces a coupling between
the inter- and intra-subband spin-density excitations which
results in the avoided crossing of inter- and intra-subband
spin-density dispersions and a mixing of intra-subband particle-hole
excitations with the inter-subband continuum.
In (b), one sees that this mode coupling also prevents the
inter-subband SDE from softening when going from the
paramagnetic to the antiferromagnetic phase.
\label{fig:asymmode}}
\end{figure}

The presence of asymmetry in the double quantum well
complicates this picture somewhat.
The SDE spectrum for our asymmetric ($dU = 0.5$~meV) double
quantum well illustrates these complications and
is shown in Fig.~\ref{fig:asymmode}(a).
Most noticeably, the asymmetry mixes the intra- and
inter-subband excitations, so that even in scattering
geometries designed to measure only inter-subband
response, both intra- and inter-subband excitations will
appear.\cite{note}
This effect is seen through both an enlarged particle-hole
continuum and the presence of a damped mode in the
inter-subband spectrum corresponding to the
intra-subband SDE.
Furthermore, the asymmetry couples the intra- and
inter-subband SDEs themselves, leading to an avoided
crossing which may be seen by comparing
Fig.~\ref{fig:asymmode}(a) to Fig.~\ref{fig:symmode}(a).

This avoided crossing effectively prevents the inter-subband
SDE from completely softening on entering the
antiferromagnetic phase, as shown in Fig.~\ref{fig:asymmode}(b).
In this figure, ones sees that, as the density is tuned
toward the antiferromagnetic transition, the $q=0$
inter-subband SDE energy decreases to a finite value.
Further increase of the density into the antiferromagnetic
phase reverses this trend, and the energy of the amplitude
mode of the antiferromagnetic order parameter increases,
resulting in a cusp.

Despite the fact that the inter-subband SDE does not
completely soften, the antiferromagnetic phase appears,
demonstrating that this softening is a sufficient but not
necessary signature of the zero-field phase transition.
We expect some mode to soften at this transition, however,
and it turns out that it is the {\it intra}-subband SDE which
collapses.
Specifically, the mode coupling between inter- and intra-subband
SDEs pushes the latter mode down in energy at all
$q$, effectively reducing the group velocity of this mode.
Approaching the antiferromagnetic transition, the
$q \rightarrow 0$ group velocity of the intra-subband SDE
decreases until it vanishes at the transition point.
The resulting phase has the character of the real-space
spin density profile of the intra-subband mode, which direct
calculation reveals to be antiferromagnetic.
The character of the intra-subband excitation at small $q$
is therefore very similar to the inter-subband excitation
in the symmetric well due to the mode coupling between
the two excitations.
Thus, its softening can lead to an antiferromagnetic
transition without inconsistency.

\section{CONCLUSION}
\label{sec:conclusion}

In this paper, we have examined the origin and robustness
of a zero-field antiferromagnetic phase in double quantum
wells predicted in previous calculations.\cite{das-tam,rad-das}
Based on a simple model, we determined that such
magnetic phases are a direct consequence of the
magnetism expected at low densities in two-dimensional
systems where interaction effects dominate.
In particular, magnetic phases with either ferro- or
antiferromagnetic spin polarizations 
are possible, and a superexchange interaction between the wells
leads to a region at intermediate densities and interaction
strengths where the antiferromagnetic phase is preferred.

By performing a detailed self-consistent Hartree-Fock
calculation on a model of these double quantum well systems,
we addressed two features left out of preceding
work:\cite{rad-das} intra-subband repulsion and asymmetry
of the heterostructure.
Our results support and extend those of Ref.~\onlinecite{rad-das}.
Specifically, we found that the antiferromagnetic phase is
stable in a large region of the model phase diagram despite
the presence of intra-subband repulsion that is as strong as
the inter-subband repulsion which drives the antiferromagnetic
instability.
The intra-subband repulsion does, however, introduce
ferromagnetic phases, producing a rich phase diagram.
Note, however, that no charge ordering phases of the type
discussed in
Refs.~\onlinecite{MacDonald,Ruden,nei,Varma,Katayama,Ying,Patel,Conti,Zheng}
were observed, even though our formalism did allow for that
possibility.
In addition, both ferro- and antiferromagnetic phases persist
in the presence of asymmetry in the quantum well structure;
indeed, the phase diagram is qualitatively unaffected by its
introduction.
Asymmetry does have a strong influence on the collective
mode spectrum, though, and induces a mode-coupling between
inter- and intra-subband spin-density excitations which
prevents the latter from completely softening at the
antiferromagnetic phase transition.
Instead, the inter-subband spin-density excitation exhibits
a cusp at the transition while the $q \rightarrow 0$ group
velocity of the {\it intra}-subband excitation vanishes.
The coupling between these two modes nonetheless lends an
antiferromagnetic character to the intra-subband excitation
in the asymmetric system and enables the collapse of its
group velocity to yield the antiferromagnetic phase.

Taken together, these results strongly indicate that, if an
exchange-correlation-induced ferromagnetic transition occurs
in a single-layer system at sufficiently low density, then
the antiferromagnetic phase in a double-layer system should
also occur.
However, the issue of whether or not the ferromagnetic phase
obtains in a single two-dimensional layer is not settled.
Variational Monte Carlo calculations\cite{cep-ald,tan-cep}
show the presence of a ferromagnetic phase between the
paramagnetic and Wigner crystal phases, but Green's function
Monte Carlo computations\cite{tan-cep} find no intervening
ferromagnetism.
More recent numerical work based on the Monte Carlo
technique\cite{cep-unp,RS,Conti} once again favors the
existence of a ferromagnetic transition.
Other calculations using density-functional theory in
the local-spin-density approximation\cite{rad-das,Zheng} also
support the existence of a ferromagnetic transition at
sufficiently low density.
Thus, although a definitive demonstration of ferromagnetism
in a single, two-dimensional layer is lacking, a large body
of evidence exists which firmly supports this hypothesis.

The precise conditions under which these phases would be
observable are somewhat harder to elucidate based on the
mean-field theory presented in this paper.
The principle difficulty is that mean-field
calculations will tend to overestimate the temperatures
and densities at which exchange-correlation-induced phase
transitions occur.
The source of this difficulty lies in the neglect of
fluctuations in the theory, which play an important role
in the low-dimensional structures considered.
If one goes to densities and temperatures deep below the
critical values of these parameters, we expect that the
mean-field theory will give an accurate qualitative
picture of the phase.\cite{MW}

In sum, work on the single layer system indicates 
that its ferromagnetic state is unlikely to occur for
electron-gas density parameters smaller than $r_s \approx 10$,
a much lower density than would be indicated by the 
Hartree-Fock approximation, for which the transition to the 
ferromagnetic state occurs at $r_s \approx 2$.  
The present work suggests that the two-layer antiferromagnetic
phase, as well as two-layer ferromagnetic phases, are likely
to be present in double-layer
systems when the density per layer approaches the low value
at which the single-layer ferromagnetic instability occurs.
For the GaAs systems studied experimentally
it therefore seems unlikely that the antiferromagnetic state will
occur for densities per layer substantially larger than
$ \approx 10^{10} {\rm cm}^{-2}$.  
However, it is exceedingly difficult to estimate the 
transition density theoretically and one must rely on experiment.

Currently, a single experimental publication regarding a
search for the antiferromagnetic phase has appeared in the
literature,\cite{Plaut} and the results are equivocal.
The authors of this study report inelastic light scattering
measurements of the long-wavelength inter-subband collective
spin-density excitations as a function of density in a double
quantum well which was expected, on the basis of 
the original theoretical work, to show the antiferromagnetic
instability.\cite{Plaut}
Instead of completely softening at a finite density, as
predicted in earlier work,\cite{das-tam,rad-das} the
inter-subband SDE shows no dramatic structure down to the
lowest densities measured.\cite{Plaut}
These results could be accounted for in at least two different
ways.  The most likely explanation is that the 
electron density of the sample, of order
$10^{11}/{\rm cm}^2$,\cite{Plaut} is above the critical
density for the antiferromagnetic transition.
Alternatively, the calculations in this paper demonstrate
that slight asymmetry in the double quantum well will
prevent a complete softening of the inter-subband SDE and
yield a cusp as a function of density.
Since the energy of the amplitude mode in the
antiferromagnetic phase is similar to that of the
inter-subband SDE away from the critical density
[cf. Fig.~\ref{fig:asymmode}(b)], measurements at a closely
spaced grid of densities may be required to detect this cusp.
In addition, the cusp may be broadened by impurity or
fluctuation effects which are beyond our mean-field
theory, further increasing the difficulty of detecting the
transition.  Thus, the current experimental results cannot exclude the
existence of the zero-field antiferromagnetic state, and its
robustness as demonstrated by the calculations in this paper
leaves us optimistic that such a unique exchange-driven phase
can occur in nature.

\section*{ACKNOWLEDGMENTS}

RJR would like to acknowledge stimulating discussions
with H.~Ehrenreich regarding the superexchange mechanism
in this system.
The work of RJR and SDS work was supported by the US-ARO
and the US-ONR.
The work of AHM was supported by the 
National Science Foundation under grant DMR-9416906.




\begin{references}
\bibitem{note}We shall continue to refer to the SDEs as
inter- and intra-subband in an asymmetric double quantum
well even though that distinction does not strictly apply.
The reader should understand these terms to refer to
the dominant character of the collective excitation with
respect to the symmetric structure.
\end{references}
\end{document}